\documentclass[aps,prl,twocolumn,groupedaddress,nofootinbib,showpacs]{revtex4}
\usepackage[dvips]{graphicx}
\usepackage{color}
\usepackage{cancel}
\usepackage{amsfonts}
\usepackage{dsfont}
\newif\ifdraft
\drafttrue
\newif\ifpreprint
\preprinttrue

\def\eqn#1{eq.~({\ref{#1}})}

\def\spa#1.#2{\left\langle#1\,#2\right\rangle}
\def\spb#1.#2{\left[#1\,#2\right]}

\newcommand{\eq}{\begin{equation}}
\newcommand{\eqe}{\end{equation}}
\newcommand{\eqa}{\begin{eqnarray}}
\newcommand{\eqae}{\end{eqnarray}}

\newbox\charbox
\newbox\slabox
\def\s#1{{      
        \setbox\charbox=\hbox{$#1$}
        \setbox\slabox=\hbox{$/$}
        \dimen\charbox=\ht\slabox
        \advance\dimen\charbox by -\dp\slabox
        \advance\dimen\charbox by -\ht\charbox
        \advance\dimen\charbox by \dp\charbox
        \divide\dimen\charbox by 2
        \raise-\dimen\charbox\hbox to \wd\charbox{\hss/\hss}
        \llap{$#1$}
}}

\def\no{\nonumber}
\def\cN{{\cal N}}

\begin{document}

\title{
\ifpreprint
\hbox{\rm \small
UUITP-29/15
$\null\hskip 13cm\null$ \hfill 
NORDITA-2015-181
\break}
\fi
\vskip 0.2cm
Complete construction of magical, symmetric and homogeneous  \\${\cal N}=2$ supergravities as double copies of gauge theories}

\author{M.~Chiodaroli,${}^{a}$ M.~G\"{u}naydin,${}^{b}$ H.~Johansson,${}^{c,d}$ and R.~Roiban${}^{b}$}

\affiliation{\vskip 0.2cm
${}^a$Max-Planck-Institut f\"ur Gravitationsphysik, Albert-Einstein-Institut, Am M\"uhlenberg 1, 14476 Potsdam, Germany\\
${}^b$Institute for Gravitation and the Cosmos, The Pennsylvania State University, University Park PA 16802, USA \\
${}^c$Department of Physics and Astronomy, Uppsala University, 75108 Uppsala, Sweden\\
${}^d$Nordita, KTH Royal Institute of Technology and Stockholm University, Roslagstullsbacken 23, 10691 Stockholm, Sweden}

\begin{abstract} 
We show that scattering amplitudes in magical, symmetric or homogeneous ${\cal N}=2$ 
Maxwell-Einstein supergravities  can be obtained as double copies 
of two gauge theories, using the framework of color/kinematics duality. The left-hand copy is ${\cal N}=2$ 
super-Yang-Mills theory coupled to a hypermultiplet, whereas the right-hand copy is a non-supersymmetric
theory that can be identified as the dimensional reduction of a $D$-dimensional Yang-Mills theory  
coupled to $P$ fermions. For generic $D$ and $P$, the double copy gives homogeneous supergravities. 
For $P=1$ and $D=7,8,10,14$, it gives the magical supergravities.  
We compute explicit  amplitudes, discuss their soft limits and study the UV-behavior at one loop. 
\end{abstract}

 \pacs{04.65.+e, 11.15.Bt, 11.30.Pb, 11.55.Bq \hspace{1cm}}

\maketitle

Perturbative calculations in gravity and gauge theory have long been considered to be on fundamentally different footing.
Gravity is characterized by a non-polynomial, 
non-renormalizable action that produces an infinite number of interaction vertices, whereas renormalizable gauge theories only have cubic and quartic interactions.  Despite these obvious differences, modern work has clarified that the perturbative expansion of gravity is directly related to that of a pair of gauge theories through a double-copy structure. 

It has long been known that the asymptotic states of gravity can be obtained as tensor products 
of gauge-theory states. 
That such a simple relationship can be extended 
to certain interacting theories was first shown 30 years ago by Kawai, Lewellen, and Tye~\cite{KLT} using  string theory.
Modern understanding of this double-copy structure
comes from work by
Bern, Carrasco, and Johansson \cite{BCJ,BCJ2}, who found a framework that  
is applicable to loop-level amplitudes and to a broader range of theories. 
The key observation is that gauge-theory amplitudes can be organized to 
expose a kinematic Lie algebra which mirrors the gauge-group color structure.
Once gauge-theory amplitudes exhibit this duality between color and kinematics, 
gravity amplitudes are obtained by substituting the color factors with equivalent kinematic objects.
This procedure doubles the kinematic structures and thus
expresses spin-$2$ theories as double copies of spin-$1$ theories~\cite{BCJ}.

The double-copy construction has proven itself to be a powerful computational tool. It fostered rapid progress in ultraviolet (UV) studies up to four loops in maximal, half-maximal and ${\cal N}=5$ supergravities~\cite{Bern:2012uf,Bern:2014sna,Bern:2013uka}.
Moreover, a class of 
black-hole solutions has been shown to exhibit a double-copy structure which 
relates them to solutions of Maxwell's equations with sources~\cite{Monteiro:2014cda, Luna:2015paa, Ridgway:2015fdl}.

The double copy permits the construction of a broad range of gravitational theories by varying the content of matter 
(spin $\le 1/2$) fields and their representations and interactions in the two gauge theories. 
Pure and matter-coupled gravities, including 
examples of Maxwell-Einstein and Yang-Mills-Einstein theories, 
are some of the theories that admit an elegant perturbative formulation 
in this framework~\cite{KLT,BCJ,BCJ2,Carrasco:2012ca,Bern:2013qca,Nohle:2013bfa,Chiodaroli:2013upa,Johansson:2014zca,Johansson:2015oia,Chiodaroli:2014xia,Chiodaroli:2015rdg}.

A systematic classification of ${\cal N}<4$ supergravities that admit double-copy constructions has not yet been obtained. 
There is a rich space of such theories, and it is not {\it a priori} obvious that the double copy can reproduce this abundance.
Indeed, in this context it is natural to ask whether the double-copy structure can be a general property of gravitational theories.

In this Letter we consider ${\cal N}=2$ Maxwell-Einstein supergravity (MESG) theories dimensionally reduced from five to four spacetime dimensions. 
These theories provide a tractable arena in which structures underlying generic gravitational theories can be uncovered. 
Unlike more supersymmetric theories, they are not uniquely specified by their matter content alone. 
However, due to their five-dimensional origin, theories in this class can
be identified from  their three-point interactions \cite{Gunaydin:1983bi}. 
Using this property, we provide a double-copy construction for three complete classes of ${\cal N}\!=\!2$ MESG 
theories: magical, symmetric, and  homogeneous theories (the latter class containing the former).
General homogeneous theories cannot be constructed as truncations of $\cN =8$ or matter-coupled $\cN=4$ supergravity; 
 their string-theory origin is also unclear. Our construction represents a major advance towards unraveling the double-copy structure of general gravitational theories. Homogeneous supergravities now constitute the largest known family of
double-copy-constructible theories.

\smallskip \noindent
{\bf Homogeneous ${\cal N}\!=\!2$ MESG theories:}
While we are ultimately interested in  MESG theories in four dimensions, 
we shall begin our analysis in five dimensions. Unlike $4D$ theories, the  full U-duality groups of 
$5D$, ${\cal N}=2$ MESG theories are symmetries of their Lagrangians. Furthermore, ${\cal N}=2$ MESG theories 
that describe low-energy effective theories of compactified M/superstring theory admit uplifts to five dimensions once quantum corrections are neglected \cite{Aspinwall:2000fd}.
When coupled to ${n}$ 
vector multiplets, such five-dimensional theories contain 
$(n+1)$ abelian vector fields
$A_\mu^I \, (I,J=0,\ldots, n)$, $n$ real scalar fields $\phi^x$ $(x,y=1,\ldots ,n)$, and $n$ symplectic-Majorana spinors. 
Their Lagrangian is \cite{Gunaydin:1983bi}:
\begin{eqnarray}
e^{-1}\mathcal{L}&=&-\frac{1}{2}R- \frac{1}{4}
{\stackrel{\circ}{a}}_{IJ} F_{\mu\nu}^{I}F^{\mu\nu J}- \frac{1}{2}
g_{xy}(\partial_{\mu}\phi^{x})(\partial^{\mu}
\phi^{y}) \no \\
&& +\frac{e^{-1}}{6\sqrt{6}}C_{IJK}
\varepsilon^{\mu\nu\rho\sigma\lambda}F_{\mu\nu}^{I}
F_{\rho\sigma}^{J}A_{\lambda}^{K} + \text{fermions} \ , \quad \label{Lsugra}
\end{eqnarray}
where $F^I_{\mu \nu}$ are abelian field-strengths.  A remarkable  property of these theories is that 
the Lagrangian is uniquely determined by the constant symmetric tensor $C_{IJK}$ whose invariance group coincides with the U-duality group. 
The scalar manifold of $5D$ MESG theories  can be interpreted as the hypersurface defined by 
$\mathcal{V}(\xi)\equiv (2/3)^{3/2}C_{IJK}\xi^{I} \xi^{J} \xi^{K} = 1$ in an $({n}+1)$-dimensional ambient space with   the metric  
\begin{equation}\label{aij}
a_{IJ}(\xi)\equiv -\frac{1}{2}\frac{\partial}{\partial \xi^{I}}
\frac{\partial}{\partial \xi^{J}} \ln \mathcal{V}(\xi)\, .
\end{equation}
The matrix ${\stackrel{\circ}{a}}_{IJ}$ in the kinetic-energy term of the vector fields
is the restriction of the ambient-space metric 
to the constraint surface, while the metric $g_{xy}$ of the scalar manifold is the pullback of the
ambient-space metric to the constraint surface.

The given structure is sufficient to calculate the bosonic part of the amplitudes we will discuss
in this letter.  We refer the reader to  \cite{Gunaydin:1983bi} for further details and  
for the fermionic terms.  These terms involve a symmetric tensor $T_{xyz}$ 
which is the pullback of the $C$-tensor to the constraint surface.  
The $C$-tensors of the theories with covariantly-constant  $T_{xyz}$ are defined by the cubic norms of Euclidean 
Jordan algebras of degree three, and their scalar manifolds are symmetric spaces \cite{Gunaydin:1983bi}.
The four magical MESG theories are defined by simple Jordan algebras 
of Hermitian $3\times 3$ matrices over  real and complex numbers, quaternions, and octonions 
 \cite{Gunaydin:1983rk}. They are the only {\it unified}  MESG   theories in five dimensions with homogeneous scalar manifolds. The unique unified $5D$ Yang-Mills-Einstein theory with homogeneous scalar manifold is obtained by gauging the quaternionic magical MESG theory~\cite{Gunaydin:1984nt}.  
The  generic Jordan theories are defined by the infinite family of non-simple Jordan algebras of degree three. 
These two classes  exhaust the list of $5D$ MESG theories with symmetric target spaces such that the full isometry group is a symmetry of the Lagrangian. There exists another infinite family of MESG theories with symmetric target manifolds, the so-called generic non-Jordan family~\cite{Gunaydin:1986fg}, where not all the isometries of the target manifolds extend to symmetries of the $5D$ Lagrangian~\cite{deWit:1991nm}.  

The most general form of the $C$-tensor consistent with unitarity was given in ref.~\cite{Gunaydin:1983bi} and depends on $n ( n^2-1)$ parameters. 
The cubic norms ${\cal V}(\xi)$ of MESG theories with homogeneous scalar manifolds and a transitive group of isometries can be brought to the form 
\cite{deWit:1991nm} 
\begin{equation} 
{\cal V}(\xi) = \sqrt{2}\big( \xi^0 (\xi^1)^2 - \xi^0 (\xi^i)^2 \big) + \xi^1 (\xi^\alpha)^2  + \tilde \Gamma^i_{\alpha \beta} \xi^i \xi^\alpha \xi^\beta  \ ,
\label{HomForm}
\end{equation}
where $i,j = 2,3, \ldots, q+2$ and $\alpha, \beta$ are indices with range~$r$.
$\tilde \Gamma^i_{\alpha \beta}$ are symmetric gamma matrices forming a real representation of the Clifford algebra ${\cal C}(q+1,0)$. 
${\cal V}(\xi)$ in eq.~(\ref{HomForm}) are generically labeled by two integers $q \geq -1$ and $P \geq 0$, except when 
$q=0,4$ (mod $8$), in which case the extra parameter $\dot P \geq 0$ is also present.  

The corresponding MESG theories give the coupling of $n=(2 + q + r)$  vector multiplets to the gravity multiplet in $5D$,  with $r = P {\cal D}_q$
or $r = (P + \dot P){\cal D}_q$.
The values for  ${\cal D}_q$ are listed in table \ref{tab1}.
The generic Jordan family corresponds to $q=\dot P=0$ and $P$ arbitrary and to $P=\dot P= 0$ and $q$ arbitrary; 
the magical theories correspond to $P=1$ and $q=1,2,4,8$, while the generic non-Jordan family theories correspond to $q=-1$.

Upon dimensional reduction the Lagrangian~(\ref{Lsugra}) can be used to describe four-dimensional MESG theories. The homogeneous $\mathcal{N}=2$
MESG theories in $4D$ were first classified in ref.~\cite{Cecotti:1988ad} using the so-called C-map and the known classification of homogeneous 
quaternionic manifolds~\cite{alekseevskii}. However, the list of ref.~\cite{Cecotti:1988ad} is not complete.
There exists an additional infinite family of homogeneous theories that descend from the generic non-Jordan family, 
which under the C-map lead to a novel family of homogeneous quaternionic manifolds~\cite{deWit:1991nm}. A further infinite family of $4D$ MESG theories can be found with symmetric target manifolds $SU(n,1)/U(n)$, which does not descend directly from $5D$ \cite{Cremmer:1984hc} but that can be obtained by truncation from the generic Jordan family.

The bosonic spectrum of the 4D MESG theory that descends from $5D$  contains the graviton, $(n + 2)$ vectors $A^{-1}_\mu, \ldots, A^{n}_\mu $
and $(n +1)$ complex scalars $z^0,\ldots , z^{n}$.
The $4D$  Lagrangian    
 is associated to the following  holomorphic prepotential in a symplectic formulation \cite{deWit:1984wbb,Cremmer:1984hj,deWit:1984rvr,Cecotti:1988ad},
\begin{equation} F(Z^A) = - {2 \over 3 \sqrt{3}} {C_{IJK} Z^I Z^J Z^K \over Z^{-1}} \ ,
\end{equation}
where $Z^A(z)$ are holomorphic functions of the scalars $z^I$. To carry out perturbation theory 
it is necessary to expand the Lagrangian around some base point, for which a  
canonical choice is $Z^A =  \big(1,  {i \over 2}, {i \over \sqrt{2}}, 0, \ldots , 0 )$.
We then  redefine (and dualize) 
fields to enlarge  the  manifest symmetry and obtain canonically-normalized quadratic terms.
We refer the reader to the supplemental material \cite{suppl} and to ref.  \cite{Chiodaroli:2015rdg} for technical details.
Finally, the resulting Lagrangian is used to construct 
amplitudes which are compared with the ones from the double copy.

\begin{table}[t]
\centering
\begin{tabular}{ccccc}
$q$ &  ${\cal D}_q$ & $r(q,P,\dot{P})$ & conditions & flavor group \\  
\hline
$-1$&  $1$ & $P$ &  R &   $SO(P)$   \\ 
$0$ & $1$ & $P \! + \! \dot P$ & RW   & $SO(P)\! \! \times \! SO(\dot{P})$  \\
$1$  &  $2$ & $2P$ & R & $SO(P)$  \\
$2$  & $4$ & $4P$ & R/W  & $U(P)$ \\
$3$ &  $8$ & $8P$ & PR  & $USp(2P)$  \\
$4$  & $8$ & $8P \! + \! 8\dot P$ & PRW   & $USp(2P)\! \! \times \! USp(2\dot{P})$  \\
$5$  & $16$ & $16P$ & PR  & $USp(2P)$  \\
$6$  & $16$ & $16P$ & R/W &   $U(P)$  \\
$k \!+\! 8$  & $16 \, {\cal D}_k$ &  $16 \, r(k,P,\dot{P}) $ & as for $k$ & as for $k$     \\
\end{tabular}
\caption{\small Parameters in the construction of homogeneous MESGs as double copies. 
The second column gives the parameter ${\cal D}_q$, the third column gives the number $r$ of $4D$ irreducible spinors
in the non-supersymmetric gauge theory,
which can obey a reality (R), pseudo-reality (PR) or Weyl (W) conditions.  
The flavor group is listed in the last column. \label{tab1}}
\vskip -15pt
\end{table}

\smallskip

{\bf Double-copy construction:}
The $m$-point amplitudes of Yang-Mills (YM) theories are naturally represented by cubic graphs
labeled by their topology, gauge-group representations of internal and external edges, and particle 
momenta. 
The $i$-th graph is associated to
the product of the corresponding propagators, to a color factor $c_i$ constructed by dressing each cubic vertex by 
the Clebsh-Gordan coefficient of the representations of the three fields (structure constants or group generators), 
and to a kinematic numerator $n_i$ encoding the remaining state dependence.
To construct an amplitude which manifestly obeys color/kinematics duality one must find kinematic numerators 
with the same symmetries and algebraic identities as the color factors~\cite{BCJ}. Schematically
\eq
\label{dualityEq}
c_i-c_j = c_k~~\Leftrightarrow~~n_i - n_j = n_k \ ,
\eqe
where the color factor identities stem from the commutation/Jacobi relations of the gauge group and thus involve three graphs.

The double-copy principle states that, once duality-satisfying numerators are found, the $L$-loop amplitudes of a supergravity theory are given by 
\eq
 \mathcal{M}^{(L)}_m \!= \!i^{L+1}\Big(\frac{\kappa}{2}\Big)^{2L+m-2} \!\!\!\! \sum_{i\in {\rm cubic}} \! \int \!
\frac{d^{LD}\ell}{(2\pi)^{LD}}\frac{1}{S_i}\frac{n_i\tilde{n}_i}{\prod_{\alpha_i} \! s_{\alpha_i}},
\label{Sugra}
\eqe
where $\kappa$ is the gravity coupling, $S_i$ are symmetry factors, and $1/s_{\alpha_i}$ are propagator denominators. The $n_i,\tilde{n}_i$ may be
identical or distinct gauge-theory numerators. 
The formula is valid if at least one of the two sets of numerators satisfies
manifestly the duality~\cite{BCJ2,Square}. In our construction, we obtain $n_i$ from a supersymmetric gauge theory, and $\tilde{n}_i$ from a non-supersymmetric one.
It is critical that we consider gauge theories 
with some fields transforming in a generic representation $R$ of the gauge group. 
Indeed, a judicious choice for the representation $R$
will enable us to capture a larger set of supergravities with our construction.

\smallskip \noindent
{\bf The gauge-theory copies:}
The first (left) gauge theory entering the construction is an $\cN=2$ super-Yang-Mills (SYM) theory with a single half-hypermultiplet 
transforming in a pseudo-real representation $R$ of a gauge group $G$.
To be precise, pseudo-real means that there exists 
a unitary antisymmetric matrix $V$ obeying $V T^{\hat a} V^\dagger = - (T^{\hat a})^*$, where $T^{\hat a}$ are the representation matrices.
With this choice, the half-hypermultiplet is $CPT$ self-conjugate and we do not need to include additional fields in the theory.
We will see that using the smallest possible multiplet allows to formulate double-copy constructions for larger classes of supergravities, including
in particular all the magical supergravity theories. 
A canonical example for $R$ is the fundamental representation of $USp(2N)$. 
Note that the matrix $V$ can be used to lower or raise gauge representation indices.

The on-shell spectrum of the supersymmetric gauge-theory factor is
\begin{displaymath}
 \big( A^{\hat a}_+ , \psi^{\hat a}_+ , \phi^{\hat a} \big)_G \oplus \big( A^{\hat a}_- , \psi^{\hat a}_- , \bar \phi^{\hat a} \big)_G \oplus 
\big( \chi_+ , \varphi_1 , \varphi_2 , \chi_- \big)_R \ , \end{displaymath}
where $\hat a, \hat b$ are adjoint indices of $G$, and indices corresponding to the representation $R$ are suppressed. 
Amplitudes in this theory can  be conveniently organized into superamplitudes 
with manifest ${\cal N}=2$ supersymmetry~\cite{suppl}. 

The second (right) gauge theory  is a non-supersymmetric YM theory with $(q+2)$ 
scalars and $r$ fermions. Its Lagrangian is
\begin{eqnarray} {\cal L} \!\!&= \!\!& - {1 \over 4} F^{\hat a}_{\mu \nu} F^{\hat a \mu \nu} + {1 \over 2} (D_\mu \phi^a)^{\hat a} (D^\mu \phi^a)^{\hat a}  \no 
+{i \over 2}  \overline{\lambda}^\alpha  D_{\mu} \gamma^\mu \lambda_\alpha \ \no \\ && \!\!\! \!\!\! \null  + {g \over 2} 
\phi^{a\hat a} \Gamma^{a \ \beta}_{\alpha} \overline{\lambda}^\alpha \gamma_5 T^{\hat a} \lambda_\beta 
 - {g^2 \over 4} f^{\hat a \hat b \hat e} f^{\hat c \hat d \hat a} \phi^{a \hat a} \phi^{b \hat b} \phi^{a \hat c} \phi^{b \hat d}
 . \ \ \label{Lfermion} \end{eqnarray}
The scalars transform in the adjoint representation of $G$, while fermions transform in the pseudo-real representation $R$.  As before $\hat a, \hat b$ are adjoint gauge-group indices, while $\alpha, \beta=1,\ldots,r$ and $a,b=1,\ldots,q+2$ are global-symmetry indices. Spacetime spinor indices and indices associated to the representation $R$ are not displayed.
Imposing color/kinematics duality on the numerators of four-point amplitudes~\cite{suppl} gives the following constraint in the two-scalar-two-fermion case:
\begin{equation}
 n_u - n_t = n_s  \quad  \rightarrow  \quad \{ \Gamma^a , \Gamma^b \} = 2 \delta^{ab} \ ,  
\end{equation}
{\it i.e.} that the constant matrices $\Gamma^a$ appearing in the Yukawa couplings 
form a $(q+2)$-dimensional Clifford algebra. 
It is convenient to think of the theory above as the dimensional reduction of a
$(q+6)$-dimensional YM theory with fermionic matter to four dimensions. 
From a higher-dimensional perspective, the spinor $\lambda_{\alpha}$ 
includes $P$ copies (or flavors) of irreducible $SO(q+5,1)$ spinors, obeying reality (R) or pseudo-reality (PR) conditions: 
\begin{equation} 
\overline{\lambda} =  \lambda^t  {\cal C}_4 C V \ , \qquad
{\rm R:} \ \ C =   {\cal C}_{q}   \ , 
 \quad {\rm PR:} \ \ C =   {\cal C}_{q}  \Omega \ , \label{RPRcond}
\end{equation}
where ${\cal C}_{q}$  and ${\cal C}_{4}$ are the $SO(q+2)$  and $SO(3,1)$ charge-conjugation matrices with 
${\cal C}_q \Gamma^a {\cal C}_q^{-1} = - \zeta (\Gamma^a)^t$, ${\cal C}_4 \gamma^\mu {\cal C}_4^{-1} = - \zeta (\gamma^\mu)^t$, $\zeta = \pm1$.
$\Omega$ is an anti-symmetric real matrix acting on the flavor indices, $V$ is the matrix in the 
pseudo-reality condition for the gauge representation matrices, and $C$ is unitary.
R conditions are appropriate for $q=0,1,2,6,7$ (mod $8$) and generically yield a $SO(P)$ manifest flavor symmetry. PR conditions are 
imposed for $q=3,4,5$ (mod $8$) and  yield a $USp(2P)$ flavor symmetry.

For even $q$, we can impose Weyl conditions of the form $\Gamma_* \lambda = \pm \lambda$, where $\Gamma_*$ is the chirality matrix.
For $q=0,4$ (mod $8$), Weyl conditions are compatible with R and PR conditions, and 
the representations with different chiralities are inequivalent. Hence the corresponding theories are parameterized by 
two distinct integers $P$ and $\dot P$ counting the number of representations of each kind.
Finally, for $q=2,6$ (mod $8$) one can rewrite the Lagrangian in terms of  Weyl spinors, 
enhancing the  manifest flavor symmetry to $U(P)$.

From a double-copy perspective, 
the resulting $4D$ supergravity theory has one vector multiplet for each $4D$ fermion in the non-supersymmetric gauge theory.
The various possibilities are listed in table \ref{tab1},  which 
provides a novel perspective on
the results of ref.~\cite{deWit:1991nm}. 
In particular,  the parameter ${\cal D}_q$ introduced in that paper
equals the minimal number of $4D$ fermions in the non-supersymmetric gauge theory. Indeed, a large part of the supergravity symmetry is already manifest in this gauge theory.  
The full U-duality Lie algebras of $4D$ homogeneous supergravity theories decompose as ${\cal G}={\cal G}_0 \oplus  {\cal G}_1 \oplus  {\cal G}_2$ with
    \begin{eqnarray}
    &&{\cal G}_0= so(1,1) \oplus so(q+2,2)\oplus 
    \mathfrak{s}_q(P,\dot{P}) \ ,  \no \\ &&
     {\cal G}_1=(1, \text{spinor}, \text{vector}), \quad   {\cal G}_2= (2,1,1)\ , \label{4dsymmetry}
    \end{eqnarray}
where  $\mathfrak{ s}_q(P,\dot{P})$ is the flavor group in table~\ref{tab1}, and the grade $1$ and $2$ generators are labeled
by their grade zero representations. 
The $4D$ supergravity theories with symmetric target spaces have additional symmetry generators corresponding to the grade $-1$ and $-2$
subspaces of the isometry Lie algebras
    \cite{Gunaydin:1983bi}.

\smallskip \noindent
{\bf Amplitudes from the double copy:}
For the $(\cN=2)\otimes (\cN=0)$ construction given here, the identification of supergravity states with the double-copy
of asymptotic gauge-theory fields is as follows:
\begin{eqnarray}
A^{-1}_- =  \bar \phi \otimes A_- \ ,    & ~~~ & h_- = A_- \otimes A_- \ , \no \\
A^0_- = \phi \otimes A_-  \ ,  & ~~~ & i \bar z^0 = A_+ \otimes A_-  \ , \no \\ 
A^a_- = A_- \otimes \phi^a  \ ,  & ~~~ &  i \bar z^a = \bar \phi \otimes \phi^a \ , \no \\
A_{\alpha -} = \chi_-  \otimes (U \lambda_-)_{\alpha}  \ ,  & ~~~ & i \bar z_{\alpha} = \chi_+ \otimes  (U\lambda_-)_{\alpha}  \ , \quad
\label{map}
\end{eqnarray}
with similar relations for the $CPT$-conjugate states. Here $U$ is a unitary rotation of the spinors in the non-supersymmetric theory.
The scalar-fermion-fermion amplitude in the non-supersymmetric theory takes the form 
${\cal A}^{(0)}_3 \big( 1 \phi^{a} , 2 \lambda_{\alpha-}, 3   \lambda_{\beta-} \big)  \! = \! - i g / \sqrt{2} \,
\langle  2 3 \rangle (\Gamma^a C^{-1})_{\alpha \beta} T^{\hat a} V^{-1}$, where $C$ is the matrix in \eqn{RPRcond}.
With the identification (\ref{map}), the double copy~(\ref{Sugra}) of the above and a vector-hyper-hyper amplitude gives, for example, the following vector-vector-scalar amplitude:
\begin{equation}
{\cal M}^{(0)}_3 \! \big( 1 A_-^{a} , 2 A_{\alpha-}, 3  \bar z_{\beta} \big) \! = \!  {\kappa \over 2 \sqrt{2}}  
\langle 1 2 \rangle^2 (U^t \Gamma^a C^{-1} U)_{\alpha \beta} \label{sugraeq0}
\end{equation}
where global spinor indices have been restricted to the subspaces of appropriate chiralities 
for $q=0,4$ (mod $8$).
The matrices $\Gamma^a$ are related to the matrices $\tilde \Gamma^i$ in the cubic norm (\ref{HomForm}) as 
$( U^t  \Gamma^a C^{-1}  U ) = \big( -\mathds{1}  ,    i \tilde \Gamma^i  \big)$. Explicit expressions for $U$ can be found for each $q$ as explained in the supplemental material \cite{suppl}.
We have verified that three-point amplitudes from the double copy match exactly the ones computed from the supergravity Lagrangian.

It is possible to confirm, without any reference to a Lagrangian, that our construction yields supergravities with 
scalar manifolds that are locally homogeneous close to the base point. Indeed, a generalization of the arguments of 
ref.~\cite{ArkaniHamed:2008gz} implies that all single soft-scalar limits of amplitudes vanish for scalars parameterizing
a homogeneous manifold. It is easy to see that this is so if the soft particle (scalar) transforms under a 
manifest symmetry.  
Since all the double-copy scalars except the dilaton-axion pair $z^0$ transform under the 
manifest $SO(q + 2)$ global symmetry, only the soft dilaton/axion limit requires a detailed analysis. 
Its vanishing implies that the double-copy theory is invariant under an additional $U(1)$ symmetry. We have verified
that this is indeed the case and that the tree-level amplitudes with field configurations with a total non-zero $U(1)$ charge 
vanish identically at four and five points.

Our construction carries over to loop-level amplitudes. As an example, we give the one-loop divergence for amplitudes between four identical matter
vectors:
\begin{eqnarray}
 {\cal M}^{(1)}_4 \big( 1 A_-^0, 2A_-^0 , 3 A^0_+ , 4A^0_+ \big) \Big|_{\text{div}} &=&  {b \over \epsilon} \Big( {10 \over 3} - {q \over 6} + {r \over 3}  \no
\Big) \ , \\
 {\cal M}^{(1)}_4 \big( 1 A_-^a, 2A_-^a , 3 A^a_+ , 4A^a_+ \big) \Big|_{\text{div}} &=&  {b \over \epsilon}  \Big( {10 \over 3} + {q \over 3} + {r \over 12} \no
\Big) \ ,
\end{eqnarray}
with $b=  2 i/ (4 \pi)^2 (\kappa/2)^4   \langle 12 \rangle^2 [34]^2$.
Interestingly, the two amplitudes have the same divergence when $r=2q$. This condition is satisfied only by the four magical theories, which are unified,
and by the so-called STU model ($q=r=0$) \cite{Gunaydin:1984ak}.

In conclusion, we have shown  that scattering amplitudes in 
homogeneous ${\cal N}=2$ supergravities -- including magical and symmetric theories -- 
can be obtained as double copies of two simple gauge theories using the framework of color/kinematics duality.
To date, this is the largest known family of double-copy-constructible theories. 
Color/kinematics duality  naturally requires the Clifford algebra structure that 
has been instrumental in the classification of homogeneous theories and provides an alternative perspective 
on these theories; in particular, the homogeneity of their target spaces manifests itself in the 
amplitudes' vanishing soft limits.
We note that it is straightforward to introduce supergravity 
hypermultiplets in our construction by adding scalars transforming in the representation $R$ to the non-supersymmetric gauge theory.
The double-copy approach is particularly well-suited for 
carrying out
loop-level computations.
The existence of a double-copy construction for such a large family of theories 
suggests  that the double-copy can play a fundamental role in general gravity theories. Generalizations of our construction to accommodate even larger classes of theories, including supergravities with a lower number of isometries and gauged R-symmetry groups, appear to be within reach.

\smallskip 

The research of MG was supported in part  under DOE Grant No: de-sc0010534. 
The research of RR was  supported in part  under DOE Grants No: de-sc0008745 and de-sc0013699.
The research of HJ is supported in part by the Swedish Research Council under grant 621--2014--5722, 
the Knut and Alice Wallenberg Foundation under grant KAW~2013.0235 (Wallenberg Academy Fellow).
The research of MC is supported by the German Research Foundation (DFG) 
through the Collaborative Research Centre ``Space-time-matter" (SFB 647-C6).


\begin{thebibliography}{99}

\bibitem{KLT} 
  H.~Kawai, D.~C.~Lewellen and S.~H.~H.~Tye,
  Nucl.\ Phys.\ B {\bf 269}, 1 (1986).
  doi:10.1016/0550-3213(86)90362-7

 \bibitem{BCJ} 
  Z.~Bern, J.~J.~M.~Carrasco and H.~Johansson,
  Phys.\ Rev.\ D {\bf 78}, 085011 (2008)
  doi:10.1103/PhysRevD.78.085011
  [arXiv:0805.3993 [hep-ph]].
 
 \bibitem{BCJ2} 
  Z.~Bern, J.~J.~M.~Carrasco and H.~Johansson,
  Phys.\ Rev.\ Lett.\  {\bf 105}, 061602 (2010)
  doi:10.1103/PhysRevLett.105.061602
  [arXiv:1004.0476 [hep-th]].

\bibitem{Bern:2012uf} 
  Z.~Bern, J.~J.~M.~Carrasco, L.~J.~Dixon, H.~Johansson and R.~Roiban,
  Phys.\ Rev.\ D {\bf 85}, 105014 (2012)
  doi:10.1103/PhysRevD.85.105014
  [arXiv:1201.5366 [hep-th]].

\bibitem{Bern:2014sna} 
  Z.~Bern, S.~Davies and T.~Dennen,
  Phys.\ Rev.\ D {\bf 90}, no. 10, 105011 (2014)
  doi:10.1103/PhysRevD.90.105011
  [arXiv:1409.3089 [hep-th]].

\bibitem{Bern:2013uka} 
  Z.~Bern, S.~Davies, T.~Dennen, A.~V.~Smirnov and V.~A.~Smirnov,
  Phys.\ Rev.\ Lett.\  {\bf 111}, no. 23, 231302 (2013)
  doi:10.1103/PhysRevLett.111.231302
  [arXiv:1309.2498 [hep-th]].

\bibitem{Monteiro:2014cda} 
  R.~Monteiro, D.~O'Connell and C.~D.~White,
  JHEP {\bf 1412}, 056 (2014)
  doi:10.1007/JHEP12(2014)056
  [arXiv:1410.0239 [hep-th]].

\bibitem{Luna:2015paa} 
  A.~Luna, R.~Monteiro, D.~O'Connell and C.~D.~White,
  Phys.\ Lett.\ B {\bf 750}, 272 (2015)
  doi:10.1016/j.physletb.2015.09.021
  [arXiv:1507.01869 [hep-th]].
  
\bibitem{Ridgway:2015fdl} 
  A.~K.~Ridgway and M.~B.~Wise,
  arXiv:1512.02243 [hep-th].

\bibitem{Carrasco:2012ca} 
  J.~J.~M.~Carrasco, M.~Chiodaroli, M.~Gunaydin and R.~Roiban,
  JHEP {\bf 1303}, 056 (2013)
  doi:10.1007/JHEP03(2013)056
  [arXiv:1212.1146 [hep-th]].
  
\bibitem{Bern:2013qca} 
  Z.~Bern, S.~Davies and T.~Dennen,
  Phys.\ Rev.\ D {\bf 88}, 065007 (2013)
  doi:10.1103/PhysRevD.88.065007
  [arXiv:1305.4876 [hep-th]].

\bibitem{Nohle:2013bfa} 
  J.~Nohle,
  Phys.\ Rev.\ D {\bf 90}, no. 2, 025020 (2014)
  doi:10.1103/PhysRevD.90.025020
  [arXiv:1309.7416 [hep-th]].

\bibitem{Chiodaroli:2013upa} 
  M.~Chiodaroli, Q.~Jin and R.~Roiban,
  JHEP {\bf 1401}, 152 (2014)
  doi:10.1007/JHEP01(2014)152
  [arXiv:1311.3600 [hep-th]].

\bibitem{Johansson:2014zca} 
  H.~Johansson and A.~Ochirov,
  JHEP {\bf 1511}, 046 (2015)
  doi:10.1007/JHEP11(2015)046
  [arXiv:1407.4772 [hep-th]].
  
  \bibitem{Johansson:2015oia} 
  H.~Johansson and A.~Ochirov,
  JHEP {\bf 1601}, 170 (2016)
  doi:10.1007/JHEP01(2016)170
  [arXiv:1507.00332 [hep-ph]].

\bibitem{Chiodaroli:2014xia} 
  M.~Chiodaroli, M.~Gunaydin, H.~Johansson and R.~Roiban,
  JHEP {\bf 1501}, 081 (2015)
  doi:10.1007/JHEP01(2015)081
  [arXiv:1408.0764 [hep-th]].

\bibitem{Chiodaroli:2015rdg} 
  M.~Chiodaroli, M.~Gunaydin, H.~Johansson and R.~Roiban,
  arXiv:1511.01740 [hep-th].
  
\bibitem{Gunaydin:1983bi} 
  M.~Gunaydin, G.~Sierra and P.~K.~Townsend,
  Nucl.\ Phys.\ B {\bf 242}, 244 (1984).
  doi:10.1016/0550-3213(84)90142-1

\bibitem{Aspinwall:2000fd} 
  P.~S.~Aspinwall,
  hep-th/0001001.
  
\bibitem{Gunaydin:1983rk} 
  M.~Gunaydin, G.~Sierra and P.~K.~Townsend,
  Phys.\ Lett.\ B {\bf 133}, 72 (1983).
  doi:10.1016/0370-2693(83)90108-9
  
 \bibitem{Gunaydin:1984nt} 
  M.~Gunaydin, G.~Sierra and P.~K.~Townsend,
  Phys.\ Rev.\ Lett.\  {\bf 53}, 322 (1984).
  doi:10.1103/PhysRevLett.53.322

\bibitem{Gunaydin:1986fg} 
  M.~Gunaydin, G.~Sierra and P.~K.~Townsend,
  Class.\ Quant.\ Grav.\  {\bf 3}, 763 (1986).
  doi:10.1088/0264-9381/3/5/007
  
\bibitem{deWit:1991nm} 
  B.~de Wit and A.~Van Proeyen,
  Commun.\ Math.\ Phys.\  {\bf 149}, 307 (1992)
  doi:10.1007/BF02097627
  [hep-th/9112027].
  
\bibitem{Cecotti:1988ad} 
  S.~Cecotti,
  Commun.\ Math.\ Phys.\  {\bf 124}, 23 (1989).
  doi:10.1007/BF01218467
  
\bibitem{alekseevskii} 
D.~V.~Alekseevskii, 
Izv. Akad. Nauk SSSR Set. Mat. {\bf 9}, 315-362 (1975); Math. USSR Izvesstija, {\bf 9}, 297-339 (1975). 
 
\bibitem{Cremmer:1984hc} 
  E.~Cremmer and A.~Van Proeyen,
  Class.\ Quant.\ Grav.\  {\bf 2}, 445 (1985).
  doi:10.1088/0264-9381/2/4/010
  
\bibitem{deWit:1984wbb} 
  B.~de Wit and A.~Van Proeyen,
  Nucl.\ Phys.\ B {\bf 245}, 89 (1984).
  doi:10.1016/0550-3213(84)90425-5
  
\bibitem{Cremmer:1984hj} 
  E.~Cremmer, C.~Kounnas, A.~Van Proeyen, J.~P.~Derendinger, S.~Ferrara, B.~de Wit and L.~Girardello,
  Nucl.\ Phys.\ B {\bf 250}, 385 (1985).
  doi:10.1016/0550-3213(85)90488-2

\bibitem{deWit:1984rvr} 
  B.~de Wit, P.~G.~Lauwers and A.~Van Proeyen,
  Nucl.\ Phys.\ B {\bf 255}, 569 (1985).
  doi:10.1016/0550-3213(85)90154-3

\bibitem{suppl} 
See   Supplemental   Material   at
http://link.aps.org/
supplemental/10.1103/PhysRevLett.117.011603
for details
on the choice of symplectic frame for the 4D supergravities,
the $\cN=2$ superamplitudes, a derivation of the color/kinematics duality constraints on the
$\cN=0$
Lagrangian
and details on the comparison of the double-copy amplitudes with the amplitudes derived from the supergravity
Lagrangian.


  \bibitem{Square} 
  Z.~Bern, T.~Dennen, Y.~t.~Huang and M.~Kiermaier,
  Phys.\ Rev.\ D {\bf 82}, 065003 (2010)
  doi:10.1103/PhysRevD.82.065003
  [arXiv:1004.0693 [hep-th]].
  
\bibitem{ArkaniHamed:2008gz} 
  N.~Arkani-Hamed, F.~Cachazo and J.~Kaplan,
  JHEP {\bf 1009}, 016 (2010)
  doi:10.1007/JHEP09(2010)016
  [arXiv:0808.1446 [hep-th]].
  
\bibitem{Gunaydin:1984ak} 
  M.~Gunaydin, G.~Sierra and P.~K.~Townsend,
  Nucl.\ Phys.\ B {\bf 253}, 573 (1985).
  doi:10.1016/0550-3213(85)90547-4

  

\end{thebibliography}
\end{document}